\documentclass[vecphys,graybox]{svmult}

\usepackage{mathptmx}       
\usepackage{helvet}         
\usepackage{courier}        
\usepackage{type1cm}        
\usepackage{makeidx}         
\usepackage{graphicx}        
\usepackage{multicol}        
\usepackage[bottom]{footmisc}

\usepackage{hyperref} 		
\hypersetup{colorlinks=true,
	linkcolor=[rgb]{1, 0.0, 0},
	citecolor=blue}
\usepackage{subeqnarray}
\usepackage{bm}

\newcommand{\calE}{\mathcal{E}}

\makeindex             

\begin{document}

\title*{A unified hyperbolic formulation for viscous  
fluids and elastoplastic solids}
\titlerunning{A unified hyperbolic formulation for continuum mechanics}
\author{Ilya Peshkov, Evgeniy Romenski and Michael Dumbser}
\institute{Ilya Peshkov \at Institut de Math\'{e}matiques de Toulouse, 
Universit\'{e} Toulouse III, F-31062 Toulouse, France, and Sobolev Institute of 
Mathematics, 4 Acad. Koptyug 
Avenue, 630090 Novosibirsk, Russia 
\email{peshenator@gmail.com}
\and Evgeniy Romenski \at Sobolev Institute of Mathematics, 4 Acad. Koptyug 
Avenue, 630090 Novosibirsk, Russia, and Novosibirsk State University, 2 
Pirogova Str., 630090 Novosibirsk, Russia \email{evrom@math.nsc.ru}
\and Michael Dumbser \at Department of Civil, Environmental and Mechanical 
Engineering, 
University of Trento, Via Mesiano 77, 38123 Trento, Italy, 
\email{michael.dumbser@unitn.it}
}
%
%
\maketitle

\abstract{We discuss a unified flow theory which in a 
single system of hyperbolic partial differential equations (PDEs) can describe 
the two main branches of continuum mechanics, fluid dynamics and solid 
dynamics. The fundamental difference from the classical  continuum models, such 
as the Navier-Stokes for example, is that the finite length scale of the 
continuum particles is not ignored but kept in the model in order to 
semi-explicitly 
describe the essence of any flows, that is the process of continuum particles 
rearrangements. 
To allow the continuum particle rearrangements, we admit the deformability of 
particle which is described by the distortion field. The ability of media to 
flow is characterized by the strain dissipation time which is a characteristic 
time necessary for a continuum particle to rearrange with one of its 
neighboring particles. It is shown that the continuum particle length scale is 
intimately connected with the dissipation time.  
The governing equations are represented by a system of first 
order hyperbolic PDEs with source terms modeling the dissipation due to 
particle rearrangements. Numerical 
examples justifying the reliability of the proposed approach are 
demonstrated.
}

\section{Introduction}
\label{sec:introduction}
This paper contains an extended abstract of the talk given at the XVI 
International Conference on Hyperbolic Problems
Theory, Numerics, Applications (HYP2016), Aachen (Germany), August 1-5, 2016. 
The talk was dedicated to the unified hyperbolic formulation of fluid and 
solid dynamics recently proposed in~\cite{HPR2016,DPRZ2016}. In particular, the 
emphasis was done on the discussion of the \textit{physical model} underlying 
the 
mathematical formulation. To emphasize how important 
such a physical interpretation of the mathematical model is, we recall that 
the  equations which 
constitute the model were proposed many years 
ago, back to 1970th, by Godunov 
and Romenski in \cite{GodRom1972,God1978} for modeling of large elastoplastic 
deformations in metals, and the equations were used until recently only in the 
solid dynamics context by several authors, e.g. 
~\cite{Rom1989,Resnyansky1995,Resnyansky2002,GavrFavr2008,BartonRom2010,Pesh2010,
favrie2009solid,Resnyansky2011constitutive,Barton2013,Ndanou2014,Pesh2015,
Boscheri2016}
to cite just a few. Moreover, similar equations and even an idea to apply them 
to 
modeling of fluids were proposed by Besseling in~\cite{Besseling1968}, but 
unfortunately it has never been appreciated in the fluid dynamics context nor 
by Besseling itself neither by the others.  Perhaps, one of the reason for that 
the 
hyperbolic Godunov-Romenski equations was not even thought to be used 
in the fluid dynamics context is an \textit{exceptional role} 
of the parabolic Navier-Stokes-Fourier (NSF) equations in the fluid dynamics. 
For 
example, it is believed that  any 
mathematical model aiming to describe viscous flows has to \textit{literally} 
coincide with the NSF equations in the diffusion regime. This should be 
understood as that the 
second order parabolic terms should appear \textit{explicitly}  in the PDEs and 
they are a 
fundamental \textit{hallmark} of the diffusion in the mathematical description. 
For instance, the 
well-known first order hyperbolic extension of the NSF equations, the 
Maxwell-Cattaneo equations
\begin{equation}\label{eqn.Cattaneo}
\dot{X}=-\frac{1}{\lambda}\left( X-X^{{\rm NSF}}\right),
\end{equation}
\textit{relax} to the NSF equations as the relaxation parameter $ 
\lambda\rightarrow0 $. Here, $ X $ is a dissipative 
quantity in the Maxwell-Cattaneo approach, while $ X^{\rm NSF} $  is the value 
of $ X $ 
obtained in the framework of the NSF theory, the upper dot denotes a time 
derivative. This, in particular, results in 
that 
some characteristic speeds of the Maxwell-Cattaneo equations 
\textit{unphysically} tend 
to infinity as $ \lambda\rightarrow0 $. From the other side, as it is shown 
in~\cite{HPR2016,DPRZ2016}, there are no physical reasons saying that the 
diffusion processes should be exclusively modeled by the second order parabolic 
equations, and a radically different first order hyperbolic description which 
is not based on the steady-state laws such as Newton's law of viscosity or 
Fourier law of heat conduction is possible.    

Perhaps, the right question in this context is that, after more than one 
hundred year 
history of successful use of the NSF theory, do we need at all another 
transport theory 
different from the classical parabolic approach? From a 
practical point of view, the 
answer is not clear yet, but from a physical viewpoint the answer is 
obviously positive. 
Indeed, the heart of the NSF equations, the Newton's law of viscosity and 
Fourier's law 
of heat conduction, are the \textit{phenomenological} laws, and thus should be 
substituted by more physically meaningful laws. We thus would like to emphasize 
an important role of the physical model in that it helped us to dare to 
propose an alternative physically-based description of viscous 
dissipation. Eventually, it is necessary to note that our hyperbolic unified 
approach is now well established after an extensive comparison with the NSF 
theory in~\cite{DPRZ2016}. Moreover, the model was recently 
extended in~\cite{HPR2016elmag} in order to include the interaction of matter 
with 
the electromagnetic field where we also provided an extensive comparison of 
the extended model with the ideal MHD and parabolic viscous resistive MHD 
equations.

\section{Physical model}
\label{sec:mental}

\begin{figure}[t]
\sidecaption
\includegraphics[scale=0.8]{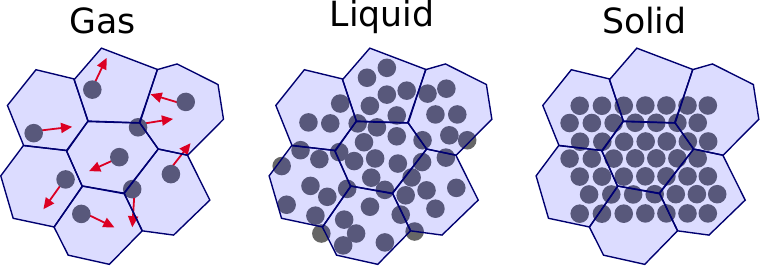}
%
%
\caption{The sketch of the continuum particles (honeycomb-like cells). If the 
scale of the continuum particles is not ignored then the continuum}
representation of all three states of matter is identical. The circles 
represent the real molecules.
\label{fig:particles}       
\end{figure}

Despite we oppose our model to the classical continuum models such as 
the 
NSF equations, we underline that the proposed approach entirely relies on the 
conventional postulates of continuum mechanics and thermodynamics. The main 
difference though is that we do not assume some simplifications which are 
implied 
in the 
classical theories. Namely, the key difference is that the \textit{continuum 
particles} 
are not treated as \textit{scaleless mathematical points} but are 
considered as the finite volumes of a small but \textit{finite scale} $ \ell $. 
Recall that the 
notion of the continuum particle is central in any continuum theory. This 
notion relies 
on the longtime observations suggesting that for the macroscopic description of 
the 
dynamics of matter (gas, liquid or solid) the very detailed information about 
the molecular motion is irrelevant but the dynamics of \textit{ensembles of 
molecules} instead should be considered as a dynamics of new entities of a 
particle nature. Of course, such particles have 
not to exist forever but only during a finite time which shall be considered 
later as an important measure of \textit{fluidity}. Thus, in our approach, 
the continuum is represented by a system of finite scale particles (finite 
volumes) covering the space occupied by the media without gaps, see 
Fig.~\ref{fig:particles}.

%

\begin{figure}[t]
\sidecaption
\includegraphics[scale=2.0]{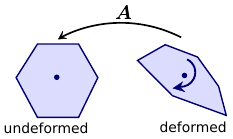}
%
%
\caption{The sketch for the distortion field $ \bm{A} $. It maps a particle 
from a current deformed state to the undeformed stress free state.}
\label{fig:dist}       
\end{figure}

Once one admits or, rather to say, does not ignore that the continuum 
particles 
have a finite scale $ \ell $, the description of the flow becomes 
straightforward because the essence of any flow phenomena in any system of 
finite scale particles is the 
process of \textit{rearrangements} of these particles\footnote{From the other 
hand, in 
the context of the scaleless particles of classical continuum mechanics it is 
impossible to define the 
rearrangements because the notion of a \textit{neighboring continuum particle} 
becomes 
\textit{indefinite}, and thus what remains is not to describe the flow itself 
but rather to \textit{mimic} some indirect flow indicators, such as 
stress--strain-rate relations, etc. Such a mimic strategy is of course 
admissible  
in the engineering problems but it is unable to give a meaningful 
explanation to the physical phenomena.}. Thus, the central task 
of our approach is \textit{to find a mathematical framework to express the 
dynamics of the continuum particles}. Further, as depicted in 
Fig.~\ref{fig:particles}, there is no \textit{free volume} between the 
continuum 
particles, and hence to allow the particle rearrangements, we have to admit the 
deformability of the particles, otherwise (if they would be rigid volumes) they 
cannot 
rearrange and the flow is impossible. Thus, in our approximation, the 
continuum particles are the structureless (homogeneous) "soft" deformable 
particles. The ability of the particles to 
deform can be characterized by the ability to transmit the \textit{transversal 
perturbations}, which in turn can be  characterized by the shear sound speed, 
denoted here by $ 
c_s $. As for the measure of the deformation of particles, we use the 
distortion 
field 
$ \bm{A}=[A_{ij}] $
which maps a particle from the current deformed state to the undeformed state, 
see 
Fig.~\ref{fig:dist}.

Furthermore, the ability of the particles to rearrange, or to change their 
neighbors, can be characterized by a time $ \tau $ which is the characteristic 
time necessary for a given particle to rearrange with one of its neighboring 
particles. Because we keep the finite scale of the continuum particles in the 
physical model, it is then obvious that such particles can not rearrange 
instantaneously because of the \textit{causality principle}, and hence the time 
$ \tau $ is also finite.
The time~$ \tau $  is a continuum interpretation of the seminal idea 
of the so-called \textit{particle settled life time} of 
Frenkel~\cite{Frenkel1955}, who applied it to describe the ability of liquids 
to flow, 
see also \cite{brazhkin2012two,bolmatov2013thermodynamic,bolmatov2015revealing} 
and  references therein. 

\begin{figure}[t]
\sidecaption
\includegraphics[scale=2]{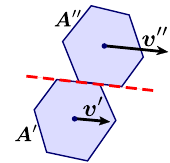}
%
%
\caption{The sketch for the particle rearrangements. Because the continuum 
particles are 
finite, there is a relative velocity $ \bm{w}=\bm{v}''-\bm{v}' $ between 
neighbouring 
particles. While the longitudinal (pressure) perturbations can propagate across 
the 
slip plane, the 
transversal perturbations can} not propagate without a loss of information, and 
thus the distortions $ \bm{A}' 
$ and $ \bm{A}'' $ are \textit{incompatible}.
\label{fig:flow}       
\end{figure}

Another nontrivial, and probably the most important, consequence of the 
finitennes of the particle length scale is that because the particles cannot 
rearrange instantaneously, there is a \textit{relative} motion between the 
neighbouring particles, see Fig.~\ref{fig:flow}. Such a relative motion assumes 
the existence of a slip between neighbouring particles. In turn, the 
transversal 
perturbations that carry the information about the deformation of the continuum 
particles cannot propagate across such slip planes without a loss of 
information. This results in that the distortion field is 
\textit{incompatible}\footnote{The incompatibility condition for $ \bm{A} $ is 
$ \bm{B}:={\rm 
curl(\bm{A})} \neq0 $, where $ \bm{B} $ is a so-called Burgers tensor which is 
interpreted as the number density of the slips (defects) between continuum 
particles. 
The term $ {\rm curl}(\bm{A}) $ 
also emerges in the time evolution for $ \bm{A} $.}, or not 
integrable~\cite{God1978,GodRom2003,HPR2016}. 
Such a loss 
of information is represented by a dissipation term in the time evolution for 
the distortion field which ``dissipates'' shear deformation stored in $ \bm{A} 
$. This term is proportional to $ 1/\tau $, and thus 
time $ \tau $ is also refereed to as the characteristic strain dissipation time 
in our 
papers~\cite{HPR2016,DPRZ2016,HPR2016elmag}.

%

%

\begin{svgraybox}
At this point, it is necessary to emphasize that the \textit{representation of 
the 
continuum by a system of finite volumes} is what actually 
\textit{unifies} all 
the three states of mater, gaseous, liquid and solid, because now, the problem 
of the continuum particle dynamics (finite volumes) is \textit{essentially a 
geometrical 
problem} (deformation problem). Such a geometrical reformulation is insensible 
to the 
\textit{content} of the 
continuum particles.
\end{svgraybox}

It is also clear from the above discussion that the main approximation of our 
physical model is the treatment of the continuum particles as structureless 
homogeneous elastic volumes. However, as it is shown 
in~\cite{HPR2016,DPRZ2016} such an approach is a 
very 
precise approximation as long as the characteristic wave length $ \lambda $ of 
the mechanical perturbations is larger than the particle length scale $ \ell $. 
Moreover, as it is proven in~\cite{DPRZ2016}, the knowledge of only the 
continuum particle dynamics is sufficient to build a unified flow theory for 
gases and liquids which incorporates the Newtonian behavior of viscous fluids 
as a particular case. On the contrary, the \textit{exceptionally different} 
molecular dynamics of gases and liquids suggests that not the molecular 
dynamics is responsible for the \textit{mathematical form} of the transport 
laws (identical in both cases) but a dynamics at a larger scale, which we 
believe is the scale of the continuum particles. Thus, the knowledge of the 
length 
scale $ \ell $ is extremely
important for understanding of the limits of applicability of our physical 
model, and as will be shown later with the dispersion 
analysis, the continuum particle length scale is of the order of $ \tau \, c_s 
$, 
i.e.
\begin{equation}\label{eqn.ell}
\ell \sim \tau \, c_s.
\end{equation}

If one needs to deal with a problem solution to which strongly depends on the 
dynamics at a scale for which $ \lambda \sim \ell $ or even $ \lambda<\ell $ 
then it is necessary to enlarge the model by providing a more accurate 
description of the perturbation propagation inside of the continuum particles.

\section{Mathematical model}

The governing PDEs are formulated for the following volume average quantities 
\begin{equation}\label{eqn:state_var}
(\bm{m},\bm{A},\rho,\sigma),
\end{equation}
where $ \bm{m}=[m_i] =\rho \bm{v}$ is the momentum density, $ \rho $ is the 
mass density, $ \bm{v}=
[v_i] $ is the velocity vector, $ \bm{A}=[A_{ij}] $ is the distortion field, 
and $ 
\sigma 
=\rho s $ is the entropy density, while $ s $ is the specific entropy. Also an 
exceptional role is played by the total energy density potential
\begin{equation}\label{eqn:total_energy}
\mathcal{E}=\mathcal{E}(\bm{m},\bm{A},\rho,\sigma)
\end{equation}
which plays the role of a generating potential as discussed in details 
in~\cite{DPRZ2016}.

The system of governing PDEs can be written as
\begin{subeqnarray}\label{eq:model_eul}
&\displaystyle\frac{\partial m_i}{\partial t}+\frac{\partial \left (m_i 
v_k + [m_l\mathcal{E}_{m_l} + \rho \calE_\rho - \calE] \delta_{ik} + 
A_{li}\calE_{A_{lk}}\right )}{\partial x_k}=0, 
\label{eq:model_eul_a}\\[2mm]
&\displaystyle\frac{\partial A_{i k}}{\partial t} + \frac{\partial (A_{il} 
v_l)}{\partial x_k} + v_j\left(\frac{\partial A_{ik}}{\partial 
x_j}-\frac{\partial A_{ij}}{\partial 
x_k}\right) = -\frac{\calE_{A_{ik}}}{\theta},\label{eq:model_eul_b}\\[2mm]
& \displaystyle\frac{\partial \rho}{\partial t}+\frac{\partial (\rho 
v_k)}{\partial 
x_k}=0,\label{eq:model_eul_cont}\\[2mm]
&\displaystyle\frac{\partial \sigma}{\partial t}+\frac{\partial (\sigma 
v_k)}{\partial 
x_k}=\frac{1}{\calE_\sigma\theta}\calE_{A_{ij}}\calE_{A_{ij}}\geq0,
\label{eq:model_eul_d}
\end{subeqnarray}
while the energy conservation law reads as
\begin{equation}\label{eq:energy_cons}
\frac{\partial \calE}{\partial t}+\frac{\partial }{\partial 
x_k}\left(\calE v_k + v_n\left ([m_l\mathcal{E}_{m_l} + \rho \calE_\rho - 
\calE]\delta_{nk} + 
A_{ln}\calE_{A_{lk}}\right )\right)=0.
\end{equation}

As in all our previous papers~\cite{HPR2016,DPRZ2016,HPR2016elmag}, the 
notations such as $ \calE_{\rho} $, $ \calE_{m_{i}} $, $ \calE_{A_{ij}} $, $ 
\calE_{\sigma} $ are used to denote the partial derivatives $ 
\partial\calE/\partial\rho $, $ 
\partial\calE/\partial m_i $, etc. Thus, to specify all the terms in the 
equations, 
that is to close the system, one needs to specify the total energy $ \calE $. 
Also, $ \theta\sim\tau $ is a relaxation parameter
which will be specified later. These two scalar functions, $ \calE $ and $ 
\theta $, are the only degrees of freedom in the model formulation. For 
example, the non-dissipative part of the PDEs, i.e. all the differential terms 
which are collected on the left-hand side, is 
\textit{complete} in the sense that no differential terms can be added  or 
removed and 
the only possibility to modify something is to change the potential $ \calE 
$. The dissipative part of the equations is the only algebraic source terms on 
the right-hand side which depend on the specification of the energy and the 
dissipation parameter $ \theta $.

The non-advective momentum flux
\begin{equation}\label{eqn:stress}
\Sigma_{ik}=-[m_l\mathcal{E}_{m_l} + \rho \calE_\rho - \calE] \delta_{ik} -
A_{li}\calE_{A_{lk}}
\end{equation}
is the stress tensor. Its form is completely defined by the structure of the 
time evolution equations while its further specification depends solely on the 
choice of the energy $ \calE $. 
Here, the scalar $ p = m_l\mathcal{E}_{m_l} + \rho \calE_\rho - \calE$ can be 
refereed to as the pressure which coincides with the classical hydrodynamic 
pressure for 
equilibrium flows. Indeed, if one introduces the specific total 
energy density $ E $ as $ \calE=\rho E $,
for which the following decomposition is usually assumed
\begin{equation}\label{eqn:E12}
E =E^{1}(\rho,s,\bm{A}) + \frac{1}{2}v_i v_i
\end{equation}
then $ p =  m_l\mathcal{E}_{m_l} + \rho \calE_\rho - \calE = \rho^2 
E^{1}_\rho$, exactly as in our previous paper~\cite{HPR2016}. The last term 
in~(\ref{eqn:stress}) represents the viscous 
stresses or elastic stresses in the case of solid dynamics.

For the further specification of the total energy potential $ \calE $, we note 
that there are three scales involved in the physical model formulation 
described in Introduction. Namely, the 
molecular scale, called here \textit{microscale}; the scale of the continuum 
particles, called here \textit{mesoscale}; and the flow scale, or observable
\textit{macroscale}. We thus assume that $ E $ is a sum of three terms each of 
which represents the amount of energy stored on the corresponding scale
\begin{equation}\label{eqn:E123}
E=E^{\rm mi}(\rho,s) + E^{\rm me}(\rho,s,\bm{A}) + E^{\rm ma}(\bm{v}).
\end{equation}
The terms $ E^{\rm mi} $ and $ E^{\rm ma} $ are conventional. They are the 
kinetic 
energy $ 
E^{\rm ma}(\bm{v}) =\frac{1}{2}v_iv_i$, which represents the part of the total 
energy 
stored in the macroscale, and an internal energy $ E^{\rm mi}(\rho,s) $ 
represents 
the kinetic energy of the molecular motion. In~\cite{DPRZ2016,HPR2016elmag}, we 
used an ideal gas equation or stiffened gas equation of state for $ E^{\rm mi}  
$ to model gases or liquids and solids, respectively

For the mesoscopic, or \textit{non-equilibrium}, part of the total energy, we 
shall use a quadratic form
\begin{equation}\label{eq:e_2}
E^{\rm me}=\frac{c_s^2}{4}G^{\rm TF}_{ij}G^{\rm TF}_{ij},
\end{equation}
where $ G_{ij}^{\rm TF} = G_{ij} - G_{ii}/3$ is the deviator of the 
tensor $G_{ij}=A_{ki}A_{kj}$,  $ c_s $ is the 
characteristic velocity of propagation of transversal perturbations, we call it 
here \textit{shear sound velocity}. In general, $ c_s $ is a function of $ \rho 
$ and $ s $.

With such a specification of the term $ E^{\rm me} $, the explicit form of the 
viscous/elastic stress (the last term in~(\ref{eqn:stress})) 
which we denote by $ \sigma_{ik} $
is
\begin{equation}\label{eqn:sigma}
\sigma_{ik} = \rho c_s^{2}G_{il} G^{\rm TF}_{lk}.
\end{equation}
The mesoscopic energy $ E^{\rm me} $ also defines~\cite{DPRZ2016} the 
dissipation 
terms as
\begin{equation}\label{eqn:dissipative_term}
\frac{\calE_{\bm{A}}}{\theta} = 
\frac{3}{\tau}|\bm{A}|^{\frac{5}{3}}\bm{A}\bm{G}^{\rm TF},
\end{equation}
where we use $ \theta = \tau c_s^2/3|\bm{A}|^\frac{5}{3}$ for $ \theta $ and $ 
|\bm{A}| $ to denote the determinant of $ \bm{A} $. In general, $ \tau $ is a 
function of the state variables $ \tau=\tau(\rho,s,\bm{A}) $ while for 
Newtonian fluids it can be taken to be constant as shown in~\cite{DPRZ2016} 
through a formal asymptotic analysis. In particular, the dependence of $ \tau 
$ on $ \bm{A} $ defines the non-Newtonian properties of fluids or controls 
the transition from elastic to plastic regime in 
solids~\cite{God1978,GodRom2003,BartonRom2010,Pesh2010}, see also numerical 
examples in the following Section~\ref{sec:num_res}.


%
\section{Numerical results}
\label{sec:num_res}

In this section, we demonstrate that the proposed model can be applied to 
modeling of nonequilibirum effects in gases as well as to modeling of viscous 
fluid flows and elastoplastic deformation in metals. 

\subsection{Non-equilibrium sound wave propagation in a viscous gas} %

\begin{figure}[t]
\sidecaption
\includegraphics[scale=0.4]{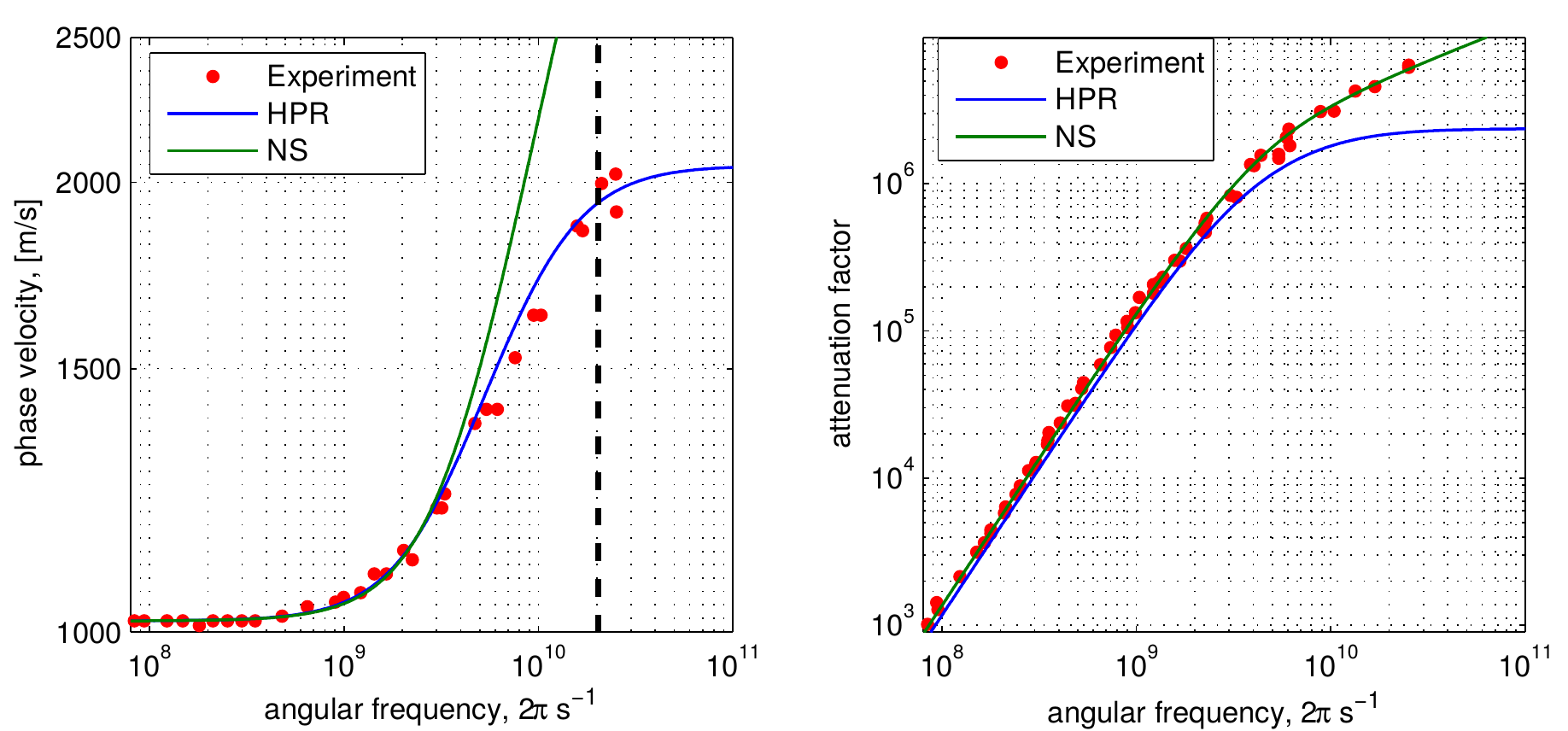}
%
%
\caption{Sound wave dispersion in Helium. Comparison with the experimental data 
(red dots) 
from~\cite{Greenspan1956}. The blue solid lines correspond to the hyperbolic 
model, and the 
green solid lines correspond to the Navier-Stokes-Fourier model. The vertical 
black} dashed line in the left figure corresponds to the frequency $\omega^* = 
2\pi/\tau  \approx 2\cdot10^{10}$ $ 2\pi{\rm \ s}^{-1} $  and to the wave 
length $ 
\lambda \approx 
10^{-7} $ 
m.
\label{fig:dispersion}       
\end{figure}

We first study the propagation of plane acoustic waves of an angular frequency 
$ \omega $ in a viscous gas. As it 
is well known the presence of the dissipative process gives rise to the 
phenomena called \textit{dispersion} when the wave phase speed $ V $  
depends on the frequency of the wave, $ V=V(\omega) $. This dependency is 
defined by the 
dispersion relation for a given model. The dispersion relation for the proposed 
hyperbolic model can be found in~\cite{HPR2016} in Section 2.2.2. The phase 
velocity $ V(\omega) $ and the attenuation factor for 
equations~(\ref{eq:model_eul}), (\ref{eqn:E123}) and (\ref{eq:e_2}) are 
presented in Fig.~\ref{fig:dispersion} for Helium.

As can be seen in Fig.\ref{fig:dispersion} (left), at a frequency $ 
\omega^*=2\pi/\tau $ the dispersion almost disappears and the phase velocity $ 
V(\omega) $ tends to a constant value $ c_\infty=\sqrt{c_0^2 + 4 c_s^2/3} $
(see also~\cite{HPR2016}) called a high frequency limit for the sound speed. 
This 
experimentally defined value can be used to estimate the shear sound speed $ 
c_s $, and subsequently to estimate the dissipation time $ \tau $ from the 
relation $ \eta = \frac{\rho}{6}c_s^2\tau$ for the shear viscosity $ \eta $.

The dispersion disappearance of $ V(\omega) $  is fully conditioned by the 
physical model 
underlying the mathematical formulation. Indeed, because the continuum 
particles have the finite scale $ \ell $, the behavior of $ V(\omega) $ should 
change when the wave length $ \lambda $ becomes comparable with the particle 
size, $ \lambda=\lambda^* \sim \ell $. We thus can use this fact to estimate 
the particle length 
scale $ \ell $ as $ \ell \sim \tau c_s $. Indeed,
\begin{equation}\label{key}
\lambda^* = 
\frac{V(\omega^*)}{\omega^*}\approx\frac{c_\infty}{\omega^*}\sim\frac{c_s}{2\pi/\tau}\sim
 \tau c_s.
\end{equation}
Thus, for the experimental data presented in Fig.\ref{fig:dispersion}, the 
continuum particle length scale can be estimated as $ \ell \approx 4.6\cdot 
10^{-7} $ m. For this, we took $ c_\infty = 2\,052 $ m/s, shear viscosity $ 
\eta 
= 
2\cdot 10^{-5} $ Pa$ \cdot $s and mass density $ \rho=0.16 $ kg/m$ ^3 $ which 
gives us $ c_s = \sqrt{(c_\infty^2 - 
c_0^2)3/4} \approx 1\,543$ m/s and $ \tau = 6 \eta /(\rho c_s^2) \approx 3\cdot 
10^{-10}$~s.

Eventually, we note that there is a certain discrepancy in the attenuation 
factor visible in the Fig.\ref{fig:dispersion}(right) if compared with the 
experimental data, while the Navier-Stokes-Fourier model (see Chapter~11 of 
\cite{EIT2010} for the dispersion relation for the NSF equations) shows an 
excellent 
agreement. First of all, one should note that, at high values of $ \omega $, 
there may be a contribution to the absorption arising from diffusion in the 
piezoelectric receiver, as pointed out by Woods and Troughton~\cite{Woods1980} 
(see also 
discussion in Chapter 11 of book~\cite{EIT2010}), so that the experimental 
result for the absorption factor should be considered as an upper limit to the 
actual value. Secondly, so far, we ignore such important processes as heat 
transfer and volume relaxation which of course should increase the dissipation. 
This will be studied elsewhere.

%

\subsection{Viscous fluids and elastoplastic solids}

In order to demonstrate the ability of the proposed unified hyperbolic model to 
deal 
with \textit{radically different} behaviors of matter such as flows of 
viscous gases and 
elastoplastic deformation in solids, we 
consider two 2D problems. These examples merely serve to demonstrate the 
diversity of 
regimes allowed to be captured by the model while the numerical 
schemes we use in this paper are not very accurate such as those used in our 
recent papers~\cite{DPRZ2016,HPR2016elmag,Boscheri2016} where much more
accurate results were obtained with the use of advanced high order ADER 
(arbitrary 
high order derivatives~\cite{Toro:2006a}) Discontinuous Galerkin and 
Finite-Volume schemes, moving mesh and adaptive mesh refinement techniques. An 
extensive comparison against the parabolic theories like the 
Navier-Stokes-Fourier equations 
and resistive MHD model is also provided in~\cite{DPRZ2016,HPR2016elmag}.

The key parameter controlling the transition between the fluid-like and 
solid-like behavior is the dissipation time $ \tau $. As discussed in the 
introduction section and in~\cite{HPR2016,DPRZ2016}, for the elastic solids, 
the continuum particles do not rearrange and hence time $ \tau $ is infinite, 
while 
for 
viscous fluids $ 0<\tau<\infty $. For elastoplastic solids, time $ \tau $ 
depends on the yield strength and rapidly but continuously changes from an 
infinite value (in fact from a sufficiently large value) to a finite 
value in the plastic regime, in which the continuum particles do rearrange.

\begin{figure}[t]
  \begin{center}
\sidecaption
\includegraphics[scale=0.3]{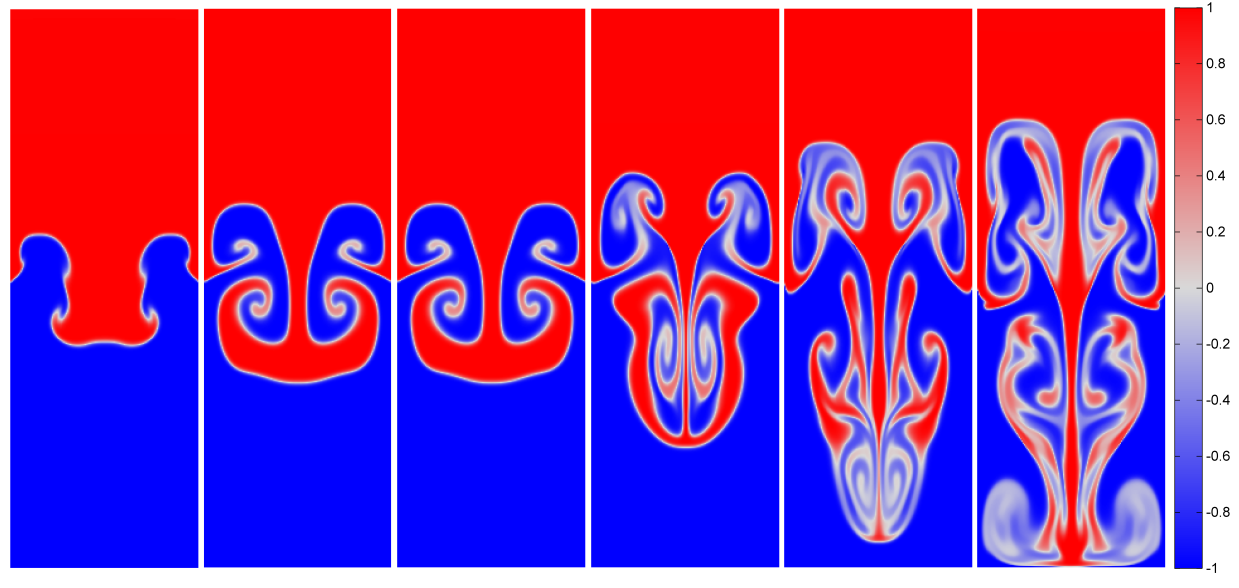}
%
%
\caption{Rayleigh-Taylor instability in a viscous gas modelled with the 
proposed hyperbolic model with the equation of 
state~(\ref{eqn:E123})-(\ref{eq:e_2}) and $ \tau={\rm const} >0$. The heavier 
gas is colored in red while the lighter gas is colored in blue. A Cartesian  
mesh of $ 
200\times 600 $ cells and no-slip boundary conditions were used.}
\label{fig:RT_instability}       
 \end{center}
\end{figure}

In the first example, a gravity driven 
Rayleigh-Taylor instability	
in a viscous gas confined in a rectangular domain with no-slip boundary 
conditions is 
simulated. The domain is a box $(x,y)\in [0,1/3]\times[0,1] $ which were 
discretized with a Cartesian mesh consisting of  $ 200\times600 $ cells, 
gravitational field is directed vertically downward and has a magnitude $ g = 
0.1 
$, the initial conditions 
are: $ \bm{v}=0 $, the density is taken 2 if $ y > 0.5+0.01 \cos(6 \pi x)$ and 
1  otherwise, the ideal gas equation of state is used for the internal energy $ 
E^{\rm mi} $ (see (\ref{eqn:E123})) with the ratio of specific heats $ 
\gamma=1.4 $ the pressure is set to  $ 1/\gamma $ everywhere, the shear sound 
speed $ c_s $ was set to $ c_s=c_0=1 $. For the whole domain, we set 
the shear viscosity to $ 10^{-5} $ Pa$ \cdot $s and the dissipation time to $ 
\tau=6\eta/ c_s^2 \approx 6\cdot10^{-5}$ s.
Fig.~\ref{fig:RT_instability} depicts several time instants of the simulation. 
The fluid-like motion (formation of the vortexes) is clearly identified. We 
also note that the gas sticks to the walls as no-slip boundary conditions is 
used.
For this simulation, we use CLAWPACK software~\cite{Clawpack} designed 
specifically for 
hyperbolic PDEs. The numerical fluxes are obtained via the solution of an 
approximate Riemann problem which was solved using the eigenvalue decomposition 
of the Jacobi matrix for the fluxes. The second order wave propagation 
algorithm of CLAWPACK with the ``minmod'' limiter was used.

\begin{figure}[!htbp]
  \begin{center}
	\begin{tabular}{c}
      \includegraphics[trim = 0mm 0mm 0mm 0mm,
      draft=false,width=0.4\textwidth]{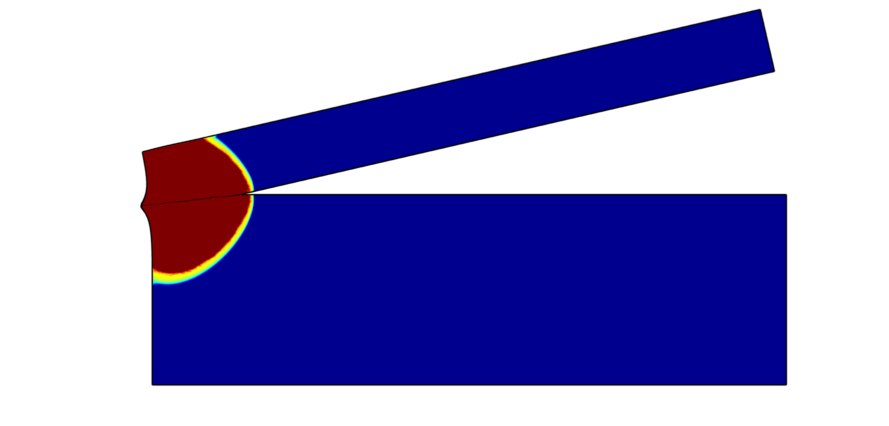} \\
      \includegraphics[trim = 0mm 0mm 0mm 0mm,
      draft=false,width=0.4\textwidth]{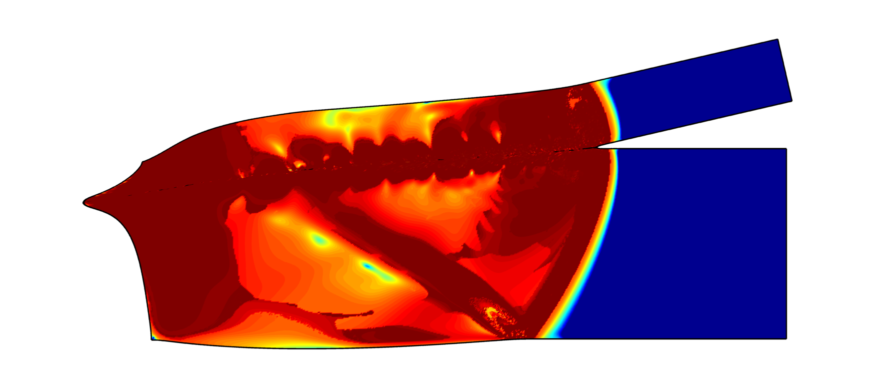} \\
      \includegraphics[trim = 0mm 0mm 0mm 0mm,
      draft=false,width=0.4\textwidth]{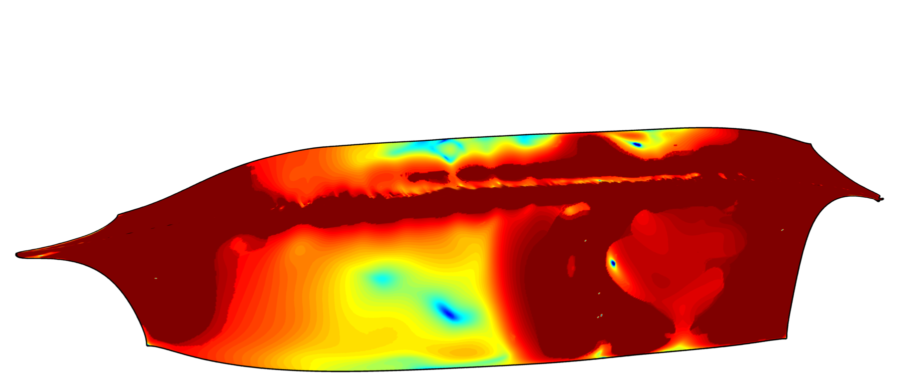}
	\end{tabular} 
    \caption{An oblique high velocity impact of two solid plates. Three 
    instants of time are shown. The colors 
    represent the 
    norm of the stress tensor deviator. The dark red corresponds to 0.04 GPa 
    and indicates the zones of plastic deformation, while all other colors 
    correspond to elastic deformation, the blue corresponds to 0~GPa. } 
    \label{fig:welding}
	\end{center}
\end{figure}

In the second numerical example, we consider an oblique high velocity collision 
of two solid 
plates presented in Fig.~\ref{fig:welding}. This example is motivated by the 
explosion 
welding process~\cite{Godunov1970welding}. The initial angle between the plates 
is 13 degrees, the lower plate is 
at rest while the upper plate has the velocity 1500 m/s normal to the bottom 
face. It is assumed that there is no slip on the contact interface between 
the bodies. The upper plate has dimensions $ 1\times0.1 $ cm and is 
discretized with a $ 600\times60 $ Cartesian mesh  while the lower 
plate has dimensions $ 1\times0.3 $ and is discretized with a $ 600\times180 $ 
Cartesian mesh.  The both plates have the same material parameters 
which were taken 
as 
follows. The mass density is 7.9 kg/m$^3$, the longitudinal sound speed is $ 
6\,700 $ m/s, the shear sound speed is $ 3\,150 $ m/s, the dissipation time was 
taken as $ \tau=\tau_0(\sigma_0/\sigma)^n $ where $ \tau_0=0.1 $~s, $ \sigma_Y 
$ is a parameter which controls the transition from elastic to plastic regime 
and was set to $ 0.0056 $ GPa, while $ \sigma $ is the second norm of the 
deviator of the stress tensor. The power law index $ n $ was set to $ n=10 $. 
For such parameters, the effective yield strength appears to be $ \approx 0.04 
$ GPa. The sound speeds were taken as $ 6\,700 $ m/s and $ 3\,150 $ m/s for 
longitudinal and transversal sound speed respectively. The equation of state is 
given in~\cite{Pesh2010}. This numerical example was performed several years 
ago, and model~(\ref{eq:model_eul}) was written in the Lagrangian coordinates 
which are well suited for the solid dynamics problems, see details 
in~\cite{Pesh2010}. 
However, recently, the Eulerian equations~(\ref{eq:model_eul})
were also implemented in an ALE (arbitrary Langrangian Eulerian) 
code~\cite{Boscheri2016} which also opens new possibilities for more 
efficient simulation of elastoplastic solids experiencing large deformations. 
Two independent Cartesian meshes were used in this simulation and the equations 
were solved with a standard first-order Godunov scheme with an acoustic Riemann 
solver, see~\cite{Pesh2010}. The contact boundary requires a specific 
treatment. For this purpose, the contact cells have to be detected and the 
numerical flux on the contact interface is obtained with the same Riemann 
solver that used for the internal cells.

%

\begin{acknowledgement}
I.P. have received funding from ANR-11-LABX-0040-CIMI within the 
program ANR-11-IDEX-0002-02 and a partial support from the Russian Foundation 
for 
Basic Research (grant number 16-31-00146). 
E.R. acknowledges a partial support by the Program N15 of the Presidium of RAS, 
project 121 and the Russian Foundation for 
Basic Research (grant number 16-29-15131). 
M.D. have received funding from the European Union's Horizon 2020 Research and 
Innovation Programme under the project \textit{ExaHyPE}, grant agreement number 
no. 671698 (call FETHPC-1-2014). 
\end{acknowledgement}
%
%
%

\bibliographystyle{vancouver}
\bibliography{library}

\begin{thebibliography}{10}

\bibitem{HPR2016}
Peshkov I, Romenski E.
\newblock {A hyperbolic model for viscous Newtonian flows}.
\newblock Continuum Mechanics and Thermodynamics. 2016;28(1-2):85--104.

\bibitem{DPRZ2016}
Dumbser M, Peshkov I, Romenski E, Zanotti O.
\newblock {High order ADER schemes for a unified first order hyperbolic
  formulation of continuum mechanics: Viscous heat-conducting fluids and
  elastic solids}.
\newblock Journal of Computational Physics. 2016;314:824--862.
\newblock Available from:
  \url{http://www.sciencedirect.com/science/article/pii/S0021999116000693}.

\bibitem{GodRom1972}
Godunov SK, Romenskii EI.
\newblock {Nonstationary equations of nonlinear elasticity theory in Eulerian
  coordinates}.
\newblock Journal of Applied Mechanics and Technical Physics.
  1972;13(6):868--884.

\bibitem{God1978}
Godunov SK.
\newblock {Elements of mechanics of continuous media}.
\newblock Nauka;.

\bibitem{Rom1989}
Romenskii EI.
\newblock {Hyperbolic equations of Maxwell's nonlinear model of elastoplastic
  heat-conducting media}.
\newblock Siberian Mathematical Journal. 1989;30(4):606--625.

\bibitem{Resnyansky1995}
Merzhievsky LA, Resnyansky AD.
\newblock {The role of numerical simulation in the study of high-velocity
  impact}.
\newblock International journal of impact engineering. 1995;17(4):559--570.

\bibitem{Resnyansky2002}
Resnyansky AD.
\newblock {DYNA-modelling of the high-velocity impact problems with a
  split-element algorithm}.
\newblock International Journal of Impact Engineering. 2002;27(7):709--727.

\bibitem{GavrFavr2008}
Gavrilyuk SL, Favrie N, Saurel R.
\newblock {Modelling wave dynamics of compressible elastic materials}.
\newblock Journal of Computational Physics. 2008;227:2941--2969.

\bibitem{BartonRom2010}
Barton PT, Drikakis D, Romenski EI.
\newblock {An Eulerian finite-volume scheme for large elastoplastic
  deformations in solids}.
\newblock International journal for numerical methods in engineering.
  2010;81(4):453--484.

\bibitem{Pesh2010}
Godunov SK, Peshkov IM.
\newblock {Thermodynamically Consistent Nonlinear Model of Elastoplastic
  Maxwell Medium}.
\newblock Computational Mathematics and Mathematical Physics.
  2010;50(8):1409--1426.

\bibitem{favrie2009solid}
Favrie N, Gavrilyuk SL, Saurel R.
\newblock {Solid--fluid diffuse interface model in cases of extreme
  deformations}.
\newblock Journal of computational physics. 2009;228(16):6037--6077.

\bibitem{Resnyansky2011constitutive}
Resnyansky AD, Bourne NK, Millett JCF, Brown EN.
\newblock {Constitutive modeling of shock response of polytetrafluoroethylene}.
\newblock Journal of Applied Physics. 2011;110(3):33530.

\bibitem{Barton2013}
Barton PT, Deiterding R, Meiron D, Pullin D.
\newblock {Eulerian adaptive finite-difference method for high-velocity impact
  and penetration problems}.
\newblock Journal of Computational Physics. 2013;240:76--99.

\bibitem{Ndanou2014}
Ndanou S, Favrie N, Gavrilyuk S.
\newblock {Criterion of hyperbolicity in hyperelasticity in the case of the
  stored energy in separable form}.
\newblock Journal of Elasticity. 2014;115(1):1--25.

\bibitem{Pesh2015}
Peshkov I, Grmela M, Romenski E.
\newblock {Irreversible mechanics and thermodynamics of two-phase continua
  experiencing stress-induced solid--fluid transitions}.
\newblock Continuum Mechanics and Thermodynamics. 2015 nov;27(6):905--940.

\bibitem{Boscheri2016}
Boscheri W, Dumbser M, Loub{\`{e}}re R.
\newblock {Cell centered direct Arbitrary-Lagrangian-Eulerian ADER-WENO finite
  volume schemes for nonlinear hyperelasticity}.
\newblock Computers {\&} Fluids. 2016 aug;134-135:111--129.
\newblock Available from:
  \url{http://linkinghub.elsevier.com/retrieve/pii/S004579301630144X}.

\bibitem{Besseling1968}
Besseling JF.
\newblock {A thermodynamic approach to rheology}.
\newblock In: Parkus H, Sedov LI, editors. Irreversible Aspects of Continuum
  Mechanics and Transfer of Physical Characteristics in Moving Fluids. IUTAM
  Symposia. Springer Vienna; 1968. p. 16--53.

\bibitem{HPR2016elmag}
Dumbser M, Peshkov I, Romenski E, Zanotti O.
\newblock {High order ADER schemes for a unified first order hyperbolic
  formulation of Newtonian continuum mechanics coupled with electro-dynamics}.
\newblock Electronic preprint. 2016 dec;Available from:
  \url{http://arxiv.org/abs/1612.02093}.

\bibitem{Frenkel1955}
Frenkel J.
\newblock {Kinetic theory of liquids}.
\newblock Dover; 1955.

\bibitem{brazhkin2012two}
Brazhkin VV, Fomin YD, Lyapin AG, Ryzhov VN, Trachenko K.
\newblock {Two liquid states of matter: A dynamic line on a phase diagram}.
\newblock Physical Review E. 2012;85(3):31203.

\bibitem{bolmatov2013thermodynamic}
Bolmatov D, Brazhkin VV, Trachenko K.
\newblock {Thermodynamic behaviour of supercritical matter}.
\newblock Nature communications. 2013;4.

\bibitem{bolmatov2015revealing}
Bolmatov D, Zhernenkov M, Zav'yalov D, Stoupin S, Cai YQ, Cunsolo A.
\newblock {Revealing the Mechanism of the Viscous-to-Elastic Crossover in
  Liquids}.
\newblock The journal of physical chemistry letters. 2015;6(15):3048--3053.

\bibitem{GodRom2003}
Godunov SK, Romenskii EI.
\newblock {Elements of continuum mechanics and conservation laws}.
\newblock Kluwer Academic/Plenum Publishers; 2003.

\bibitem{EIT2010}
Jou D, Casas-V{\'{a}}zquez J, Lebon G.
\newblock {Extended irreversible thermodynamics}.
\newblock Dordrecht: Springer Berlin Heidelberg; 2010.
\newblock Available from:
  \url{http://link.springer.com/10.1007/978-90-481-3074-0}.

\bibitem{Woods1980}
Woods LC, Troughton H.
\newblock {Transport processes in dilute gases over the whole range of Knudsen
  numbers. Part 2. Ultrasonic sound waves}.
\newblock Journal of Fluid Mechanics. 1980 sep;100(02):321--331.
\newblock Available from:
  \url{http://www.journals.cambridge.org/abstract{\_}S0022112080001176}.

\bibitem{Greenspan1956}
Greenspan M.
\newblock {Propagation of sound in five monatomic gases}.
\newblock The journal of the acousticcal society of America.
  1956;28(4):644--648.

\bibitem{Toro:2006a}
Toro EF, Titarev VA.
\newblock {Derivative Riemann solvers for systems of conservation laws and ADER
  methods}.
\newblock Journal of Computational Physics. 2006;212(1):150--165.

\bibitem{Clawpack}
{Clawpack Development Team}. {Clapack software}; 2014.
\newblock Available from: \url{http://www.clawpack.org}.

\bibitem{Godunov1970welding}
Godunov SK, Deribas AA, Zabrodin AV, Kozin NS.
\newblock {Hydrodynamic effects in colliding solids}.
\newblock Journal of Computational Physics. 1970;5(3):517--539.

\end{thebibliography}
\end{document}